\documentclass[mathleft,
]{an}
\usepackage[totalwidth=17cm,totalheight=24cm,vcentering,dvips,pdftex]{geometry}
\usepackage{graphicx}
\usepackage{times}
\usepackage[font={small},labelfont=bf]{caption}[2003/12/20]
\begin{document}
\Pagespan{343}{346}
\Yearpublication{2008}%
\Yearsubmission{2007}%
\Month{7}%
\Volume{329}%
\Issue{3}%
 \DOI{10.1002/asna.200710931}%

\title{Grid integration of robotic telescopes}

\author{F. Breitling\fnmsep\thanks{Corresponding author:
    {fbreitling@aip.de}}
  \and T. Granzer
  \and H. Enke
}
\titlerunning{Grid integration of robotic telescopes}
\authorrunning{F. Breitling, T. Granzer \& H. Enke}
\institute{Astrophysikalisches Institut Potsdam, An der Sternwarte 16, 
  D-14482 Potsdam, Germany}

\received{2007 Jul 11}
\accepted{2007 Nov 23}
\publonline{2008 Feb 25}


\abstract{Robotic telescopes and grid technology have made significant progress
in recent years. Both innovations offer important advantages over conventional
technologies, particularly in combination with one another. Here, we introduce
robotic telescopes used by the Astrophysical Institute Potsdam as ideal
instruments for building a robotic telescope network. We also discuss the grid
architecture and protocols facilitating the network integration that is being
developed by the German AstroGrid-D project. Finally, we present three user
interfaces employed for this purpose.}

\maketitle

\section{Introduction}
In recent years the first robotic telescopes came to operation and many of them
already routinely collect data without human interaction. A natural next step
in automated astronomy is the integration of these telescopes into a network. A
network of distributed telescopes can perform new types of observations
which cannot be accomplished with individual instruments. Important examples
include continuous long-term monitoring of objects independent of day and night
cycles and weather, rapid response to transient objects and multiwavelength
campaigns.

The Astrophysical Institute Potsdam (AIP) currently operates five robotic
telescopes and is pursuing their integration into a network. This effort is
supported by AstroGrid-D, which is the German astronomy community grid. Here
{\it grid} refers to a distributed network of loosely coupled resources with a
common, user-friendly infrastructure. The AstroGrid-D collaboration consists of
20 people from fourteen German institutes under the leadership of the AIP. The
project has been funded by the German Ministry of Education and Research
(BMBF) for a period of three years. The network will be built on grid
technology, which provides an ideal framework for a robotic telescope network.
For example, it provides solutions for the management of Virtual
Organization\footnote{In grid computing, a Virtual Organization defines a
collaboration of users with same access rights to grid resources.}, grid
resources, computational jobs and observation, data and metadata. In addition,
it allows the immediate access to computational and storage resources for data
analysis.

\phantom{ }

\phantom{ }

\section{Robotic telescopes of the AIP}
With control of five robotic telescopes the AIP provides the necessary hardware
for the development and operation of a telescope network. The telescopes are
RoboTel, STELLA-I and II, Wolfgang and Amadeus.

\begin{itemize}
\item
RoboTel is located at the AIP. It is a 0.8\,m telescope equipped with a CCD
camera for imaging and photometry. In addition to its science core-program, 
half of the observation time is reserved for
schools and universities. The remaining observation time is dedicated to
testing of new instruments, software and methods for the STELLA-I and II
telescopes.

\item
The STELLA robotic observatory is located at Iza{\~n}a observatory in Tenerife, Spain.
It consists of two 1.2\,m telescopes, STELLA-I and STELLA-II. STELLA-I is
equipped with a spectrograph and has been operating since May 2006. STELLA-II
will be equipped with an imaging photometer. Its commissioning starts at the
end of 2007. Scientific objectives are: Doppler imaging, the search for
extrasolar planets, spectroscopic surveys and support observations for
simultaneous observations with
larger facilities.

\item
Wolfgang and Amadeus (Strassmeier et al. 1997) 
are located at the Fairborn Observatory in Arizona. They
are two 0.75\,m telescopes equipped with photomultipliers for photometry. 
The scientific
objectives are the participation in multi-site observing campaigns and studies of
variability timescales and life times of starspots, requiring monitoring of
stars over periods of years.
\end{itemize}

Further details regarding RoboTel and STELLA can be found in Granzer (2006) and
Strassmeier et al. (2004).

\phantom{ }

\phantom{ }

\section{Grid architecture}

\begin{figure*}
  \centering
  \includegraphics[width=.66\textwidth]{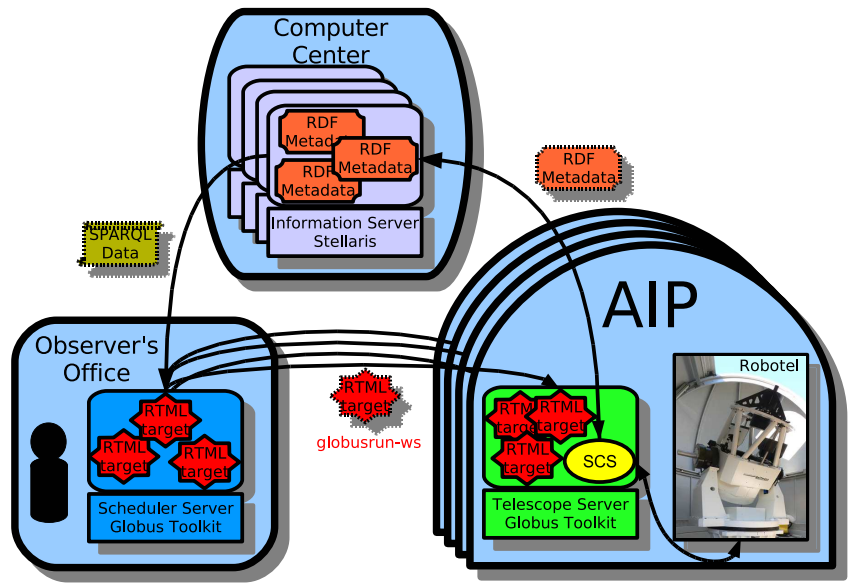}
  \caption{Grid architecture of the robotic telescope network. Telescope
  servers, scheduler and the information service ``Stellaris'' are the main
  components. The management of observation requests is done through the Globus
  Toolkit and using RTML. Metadata is submitted in RDF to Stellaris and
  retrieved using the SPARQL query language.}
  \label{fig:Architecture}
\end{figure*}

The grid architecture of AstroGrid-D consists of three main components which
are linked through the internet: the telescope servers, a scheduler and the
information service as shown in Fig.~\ref{fig:Architecture}.

The telescope server is a computer providing a robotic control system (SCS
in Fig.~\ref{fig:Architecture}) for a telescope. Therefore it is located at the
observatory. The scheduler compiles schedules for network observations, taking
into account the monitoring information on the status of each individual
telescope (including weather and engineering data). Some aspects regarding
scheduling are discussed in Granzer et al. (2007). The information service
stores and provides metadata of the entire telescope network. It plays a
central role for monitoring, scheduling and data access. It can be run at a
computer center.

The submission of observation requests and access to new data is
provided by the grid middleware: AstroGrid-D uses the Globus Toolkit (Foster
2006) commands (such as {\tt globusrun-ws}) of the Grid Resource and Allocation
Manager (WS GRAM) . The AstroGrid-D information service is called ``Stellaris''
(H\"ogqvist, R\"oblitz \& Reinefeld 2007). It is a development of the project.

\begin{figure}
  \centering
  \scalebox{.22}[.674]{\includegraphics*[1140,1602][2200,1910]{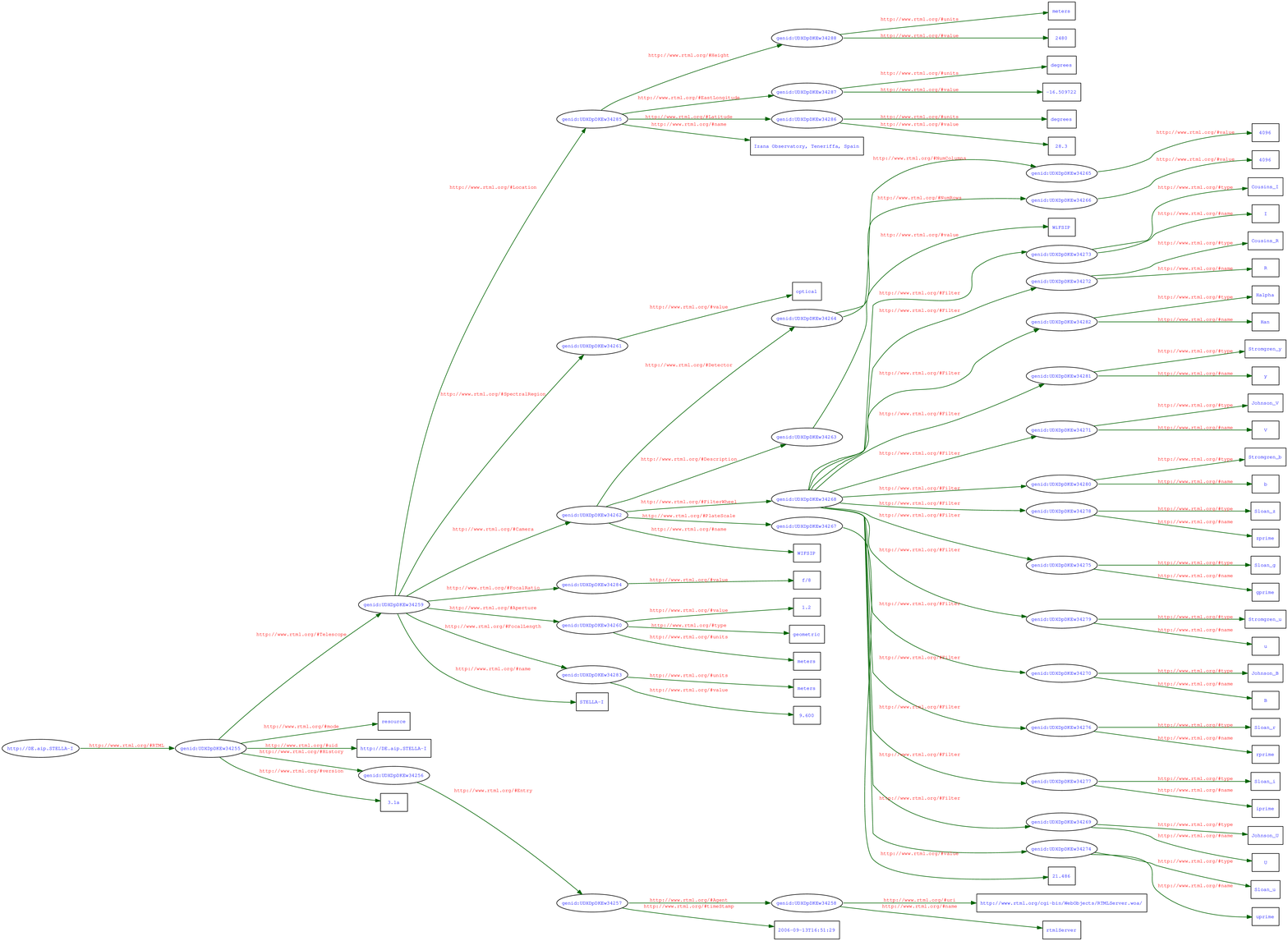}}
  \caption{Partial RDF graph of the static metadata of the robotic telescope
  STELLA-I. The represented region shows the information regarding the
  telescope's location.}
  \label{fig:RDFGraph}
\end{figure}

For the information exchange the AstroGrid-D architecture uses two protocols:
the ``Remote Telescope Markup Language'' (RTML, Hessman 2006) and the
``Resource Description Format'' (RDF 2007). RTML is an XML format for the
description of telescope metadata, such as observation requests, observation
schedules, source catalogs, telescope hardware and weather information. RDF is
a standard for storing information. It is derived from graph theory and
represents information in triples as {\it subject}, {\it predicate}, {\it
object}. Fig.~\ref{fig:RDFGraph} shows part of a RDF graph representing the
static metadata of the STELLA-I telescope. The RDF/XML and Notation 3 formats
exist as RDF representations. Since Stellaris uses RDF exclusively, a
conversion from RTML to RDF became necessary, in order to store RTML metadata.
Such a conversion has been developed using a XSL transformation (XSLT 2007).
The XSLT (rtml2rdf.xsl 2007) is the first open source transformation of
arbitrary XML documents into RDF/XML. The retrieval of information from
Stellaris is accomplished using SPARQL (2007), the query language for RDF.

\section{User interfaces}

User interaction with the network is necessary for monitoring and controlling.
Three interfaces have been developed and are discussed below.

\begin{figure}
  \includegraphics[width=\columnwidth]{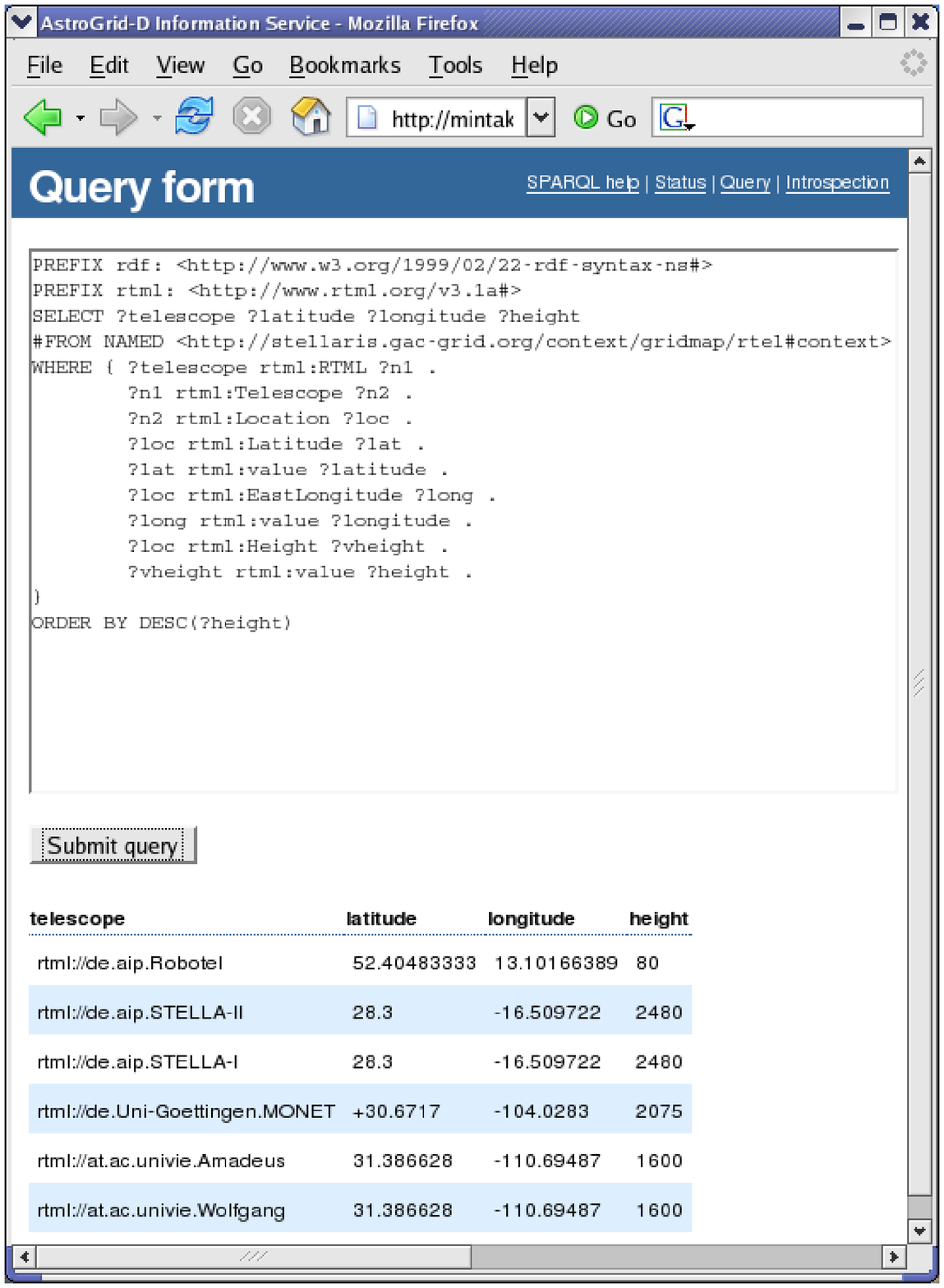}
  \caption{SPARQL query form of the web browser user interface of Stellaris.
  The query displayed here in the top section will retrieve the location of
  available telescopes sorted by altitude. The result can be seen in the lower
  half of the window.}
  \label{fig:SPARQLinterface}
\end{figure}

Stellaris provides a user interface for SPARQL queries. It is accessed through
a web browser as shown in Fig.~\ref{fig:SPARQLinterface}. If the example query
is submitted, Stellaris returns a geographic location of the available
telescopes sorted by altitude. In addition all SPARQL queries can be executed
via the command line and scripts using the ``Redland Language Bindings''
(librdf 2007).

\begin{figure*}
  \centering
  \includegraphics[width=1.6\columnwidth]{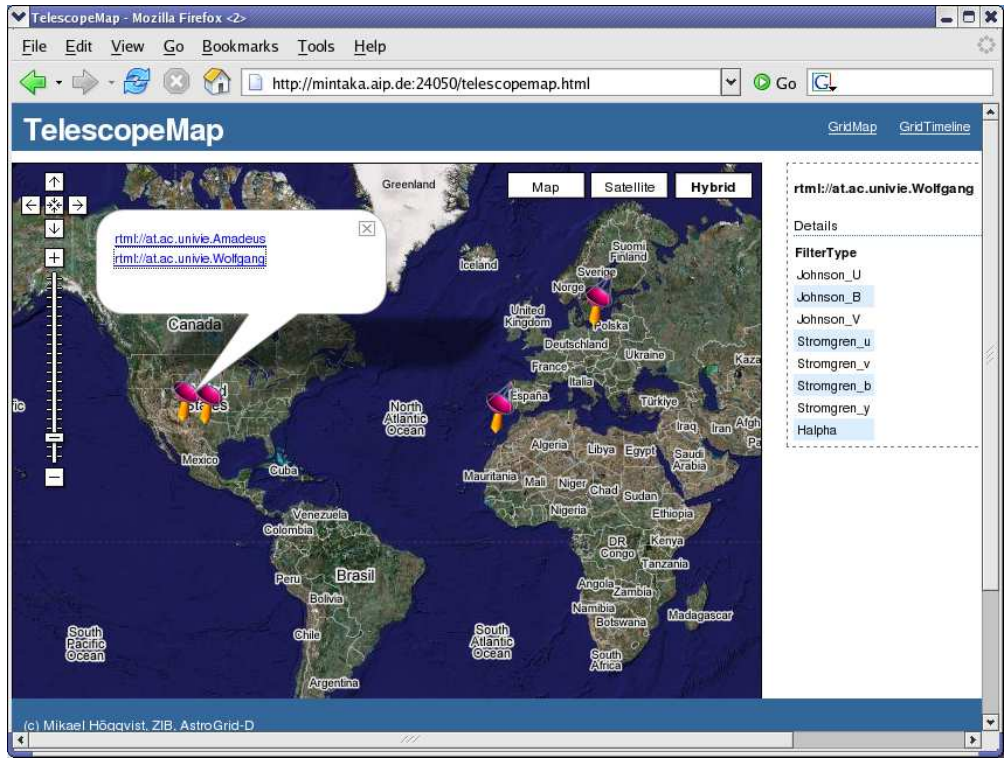}
  \caption{The Telescope Map showing the locations of observatories by bright
  telescope markers. When a marker is selected additional information about the
  installed telescopes as well as their setup is displayed.}
  \label{fig:TelescopeMap}
\end{figure*}

A second web browser user interface is the Telescope Map (Telescope Map 2007)
shown in Fig.~\ref{fig:TelescopeMap}. It is based on Google Maps (Google Maps
2007) and shows locations of observatories by markers. Selection of these
markers provides additional information such as the available telescopes and
their states. This map is useful for monitoring of network status and
observation, in particular because additional information can be added.

\begin{figure*}
  \centering
  \includegraphics[width=1.6\columnwidth]{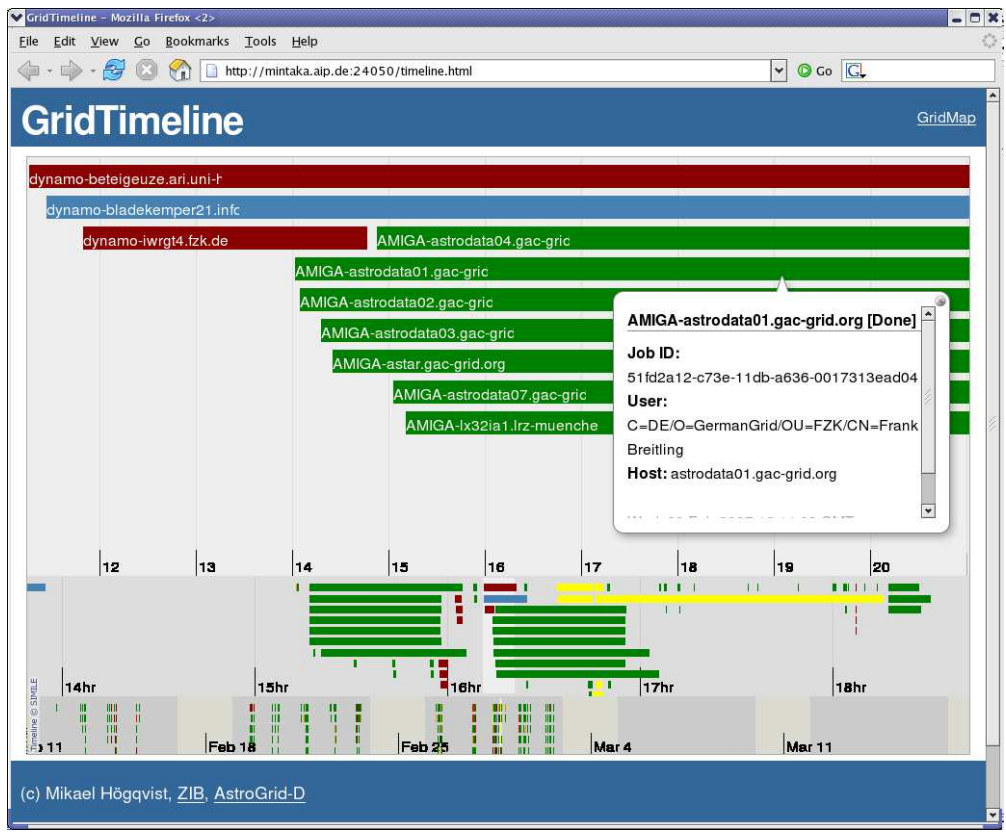}
  \caption{The Grid Timeline showing the runtime of jobs and observations
  in AstroGrid-D. Each job is represented by a horizontal bar along the
  time axis. From top to bottom three different time scales are visible.
  Additional information is displayed in a text box on selection.}
  \label{fig:Timeline}
\end{figure*}

A third web browser user interface for time-based information is the Grid
Timeline (2007) shown in Fig.~\ref{fig:Timeline}. It can be considered as
complementary to the Telescope Map. It uses the service of Simile (2007), which
provides a DHTML-based AJAXy widget for visualizing timing information. The
Grid Timeline is currently used for job monitoring and can therefore be easily
applied to monitoring of robotic observations.

\section{Conclusion}
Five robotic telescopes are operated by the AIP and are prepared for network
integration. An architecture for the integration of these telescopes to a
network using grid technology has been developed within AstroGrid-D and its
main components are available. RTML metadata of telescopes and observations can
be provided to, and received from, the information service. User interfaces are
available for easy access to this information.

\section{Outlook}
The next steps will be the implementation of the Globus job management WS GRAM
and the development of the scheduler. Then the network can undergo testing,
conduct its first observations and finally participate in campaigns with other
telescopes.

\acknowledgements The support of the German Ministry of Education and Research
(BMBF) is gratefully acknowledged as well as the support of SyncRo Soft with
the free trial license of the oXygen XML Editor (8.1.0) for Eclipse (2007).
Also the work of Mikael H\"ogqvist and Thomas R\"oblitz from the Zuse Institute
Berlin, whose contributions to the information service and user interfaces have
been very valuable, is acknowledged.


\begin{thebibliography}{}
  \bibitem{Eclipse} Eclipse: 2007, http://www.eclipse.org/
  \bibitem{Globus} Foster, I.T.: 2006, in: {\it IFIP International Conference
  on Network and Parallel Computing}, NPC 2006, p. 2
  \bibitem{Google Maps} Google Maps: 2007, http://maps.google.com/
  \bibitem{Granzer} Granzer, T.: 2006, AN 327, 792
  \bibitem{GeS2007} Granzer, T., Breitling, F., Braun, M., Enke, H.,
  R\"oblitz, T.: 2007, German e-Science Conference, http://edoc.mpg.de/316644
  \bibitem{GridTimeline} Grid Timeline: 2007,
  http://www.gac-grid.org/project-products/ Software/job-monitoring.html
  \bibitem{RTML} Hessman, F.V.: 2006, AN 327, 751
  \bibitem{Stellaris} H\"ogqvist, M., R\"oblitz, T., Reinefeld, A.: 2007,
  German e-Science Conference, http://edoc.mpg.de/316518
  \bibitem{librdf} librdf: 2007, http://librdf.org/
  \bibitem{oXygen} oXygen XML Editor: 8.1.0, http://www.oxygenxml.com/
  \bibitem{RDF} RDF: 2007, http://www.w3.org/TR/rdf-schema/
  \bibitem{rtml2rdf} rtml2rdf.xsl: 2007,
  http://www.gac-grid.org/project-products/\\ Software/XML2RDF.html
  \bibitem{Similie} Similie: 2007, http://simile.mit.edu/timeline/
  \bibitem{SPARQL} SPARQL: 2007, http://www.w3.org/TR/rdf-sparql-query/
  \bibitem{APT} Strassmeier, K.G., Boyd, L.J., Epand, D.H, Granzer, T.:
   1997, PASP 109, 697
  \bibitem{Strassmeier} Strassmeier, K.G., Granzer, T., Weber, M., et al.: 
  2004,AN325,527
  \bibitem{TelescopeMap} Telescope Map: 2007,
  http://www.gac-grid.org/project-products/ Software/TelescopeMap.html
  \bibitem{XSLT} XSLT: 2007, http://www.w3.org/TR/xslt
\end{thebibliography}
\end{document}